\documentclass[prb,twocolumn,showpacs,superscriptaddress,floatfix]{revtex4-1}
\usepackage{graphicx,amsfonts,amssymb,amsmath,hyperref}

\newif\ifhyper
\hypertrue
\ifhyper
\hypersetup{
   citecolor = {red},
   colorlinks = {true}, 
   linkcolor = {blue},
   urlcolor = {blue} 
}
\fi

\def\be{\begin{equation}}
\def\ee{\end{equation}}
\def\bea{\begin{eqnarray}}
\def\eea{\end{eqnarray}}

 \newcommand{\Zd}{\mathbb{Z}_2}
\newcommand{\ZN}{\mathbb{Z}_N}

 \newcommand{\ket}[1]{|#1\rangle}

 \newcommand{\comutator}[2]{[ #1,#2 ]}

\newcommand{\id}{1\hspace{-.25em}{\rm I}} %

\begin{document}

\title{Quantum phase transitions out of a $\mathbb{Z}_2 \times\mathbb{Z}_2$ topological phase}

\author{Saeed S. Jahromi}
\email{s.jahromi@dena.kntu.ac.ir}
\affiliation{Department of Physics, K.N. Toosi University of Technology, P.O. Box 15875-4416, Tehran, Iran}
\affiliation{Lehrstuhl f\"{u}r Theoretische Physik I, Otto-Hahn-Stra{\ss}e 4, D-44221 Dortmund, Germany}

\author{S. Farhad Masoudi}
\email{masoudi@kntu.ac.ir}
\affiliation{Department of Physics, K.N. Toosi University of Technology, P.O. Box 15875-4416, Tehran, Iran}

\author{Mehdi Kargarian}
\email{kargarian@physics.utexas.edu}
\affiliation{Department of Physics, The University of Texas at Austin, Austin, TX 78712, USA}

\author{Kai Phillip Schmidt}
\email{kai.schmidt@tu-dortmund.de}
\affiliation{Lehrstuhl f\"{u}r Theoretische Physik I, Otto-Hahn-Stra{\ss}e 4, D-44221 Dortmund, Germany}

\begin{abstract}
We investigate the low-energy spectral properties and robustness of the topological phase of color code,
which is a quantum spin model for the aim of fault-tolerant quantum computation, in the presence
of a uniform magnetic field or Ising interactions, using high-order series expansion and exact diagonalization.
In a uniform magnetic field, we find first-order phase transitions in all field directions. In contrast,
our results for the Ising interactions unveil that for strong enough Ising couplings, the $\Zd \times \Zd$ topological
phase of color code breaks down to symmetry broken phases by first- or second-order phase transitions.
\end{abstract}
\maketitle

%
%
\section{Introduction}
%
%
Topologically ordered systems are novel phases of matter which are beyond the Ginzburg-Landau symmetry-breaking theory of phase transitions \cite{landau}.
Unlike conventional phases of matter, topologically ordered states cannot be distinguished by local order parameters and their low-energy properties such as
ground-state degeneracy are characterized by non-local degrees of freedom and the topology of the Riemannian surface on which the system is embeded \cite{wen0}.
The concept of topological order was first introduced to describe
the physics of fractional quantum Hall effects \cite{tsui} and thereafter in the context of high-temperature superconductors \cite{wen1,superconductor1,superconductor2}
and frustrated magnetism \cite{RVB1,RVB2,RVB3,RVB4}.
Aside from its important role in characterizing different states of matter, topological order has a profound application in quantum computation \cite{Kitaev1}.
Highly entangled states \cite{kargarian_entaglement} and anyonic excitations in topologically ordered systems \cite{wen2,wen3} are appealing motives for building a
reliable quantum computer
by defining non-local quantum bits on the topological degrees of freedom to protect the information from local decoherence \cite{Kitaev1,Kitaev2}.

In spite of the fact that the topologically degenerate ground state of such systems is a suitable playground for quantum computation, robustness of
the topological nature of the system against local perturbations is of crucial importance. This problem has been extensively studied for the toric code which is
the simplest exactly solvable model with topological protection \cite{Kitaev1}. In the presence of a magnetic field, the $\Zd$ topologically ordered ground-state
of the toric code breaks down to a polarized phase by first- or second-order quantum phase transition according to the direction of the magnetic
field \cite{toric_terbest,toric_hamma,toric_vidal1,toric_vidal2,toric_tupitsyn,toric_dusuel,toric_wu}. The latter belongs to the 3D Ising universality
class except on a special line in parameter space where a more complicated behavior is observed \cite{toric_dusuel}. Related questions have been also
addressed for frustrated toric codes \cite{toric_kps} as well as for $\ZN$ extensions of the toric code \cite{toric_schulz}.

Another model displaying all essential features for fault-tolerant quantum computation but a different class of topological order is the so
called {\it topological color code} (TCC) \cite{bombin1_distilation}. Due to the contribution of color in the construction of the code, the topological order in this
model belongs to the $\Zd \times \Zd$ gauge symmetry group.

In this paper, we investigate the low-energy physics and robustness of the TCC, by adding a uniform magnetic field $h_\alpha$ or
ferromagnetic Ising interactions $j_\alpha$ with $\alpha=x,y,z$ to the TCC. Consequently, a large enough perturbation will destroy the topological
order of the TCC and the system has to undergo a phase transition. We find that the TCC in a magnetic field displays a first-order phase transition
between the topological phase and a polarized high-field phase for all field directions. In contrast, our results for the TCC plus Ising interactions
unveil, for the first time, a second-order quantum phase transition between a $\Zd \times \Zd$ topologically ordered phase and a $\Zd$ symmetry-broken phase
for Ising interactions $(j_x,j_z)$ while a first-order transition is found for a pure interaction $j_y$. The universality is typically 3D Ising which is shown
rigorously for a single Ising interaction $j_x$ or $j_z$ via a mapping to the transverse field Ising model (TFIM) on the dual triangular lattice. Different
critical behavior is found on a multicritical line with $j_x=j_z$ and finite $j_y$ where critical exponents appear to vary continuously which is very similar
to the behavior found for the toric code in a field \cite{toric_dusuel}. Interestingly, our results for this isotropic plane ($j_x=j_z$)
 suggest the existence of a first-order plane and a gapless phase which is adiabatically linked to the gapless $U(1)$ symmetry broken XY model in the limit of large Ising
interactions.

In order to compute the phase boundary of the topological phase, we use perturbative continuous unitary transformations
(pCUTs) \cite{pcut1,pcut2} and exact diagonalizations (ED) on periodic clusters. We compute the ground-state energy per spin as well as low-energy gaps with pCUT.
Inside the topologically ordered phase, this is done for all field directions and choices of Ising interactions. In contrast, series expansions inside the polarized
phase or the $\Zd$ symmetry-broken phase are set up only for specific parameter lines of interest.

%
%
\section{Topological Color Code}
\begin{figure}
\centerline{\includegraphics[width=8cm]{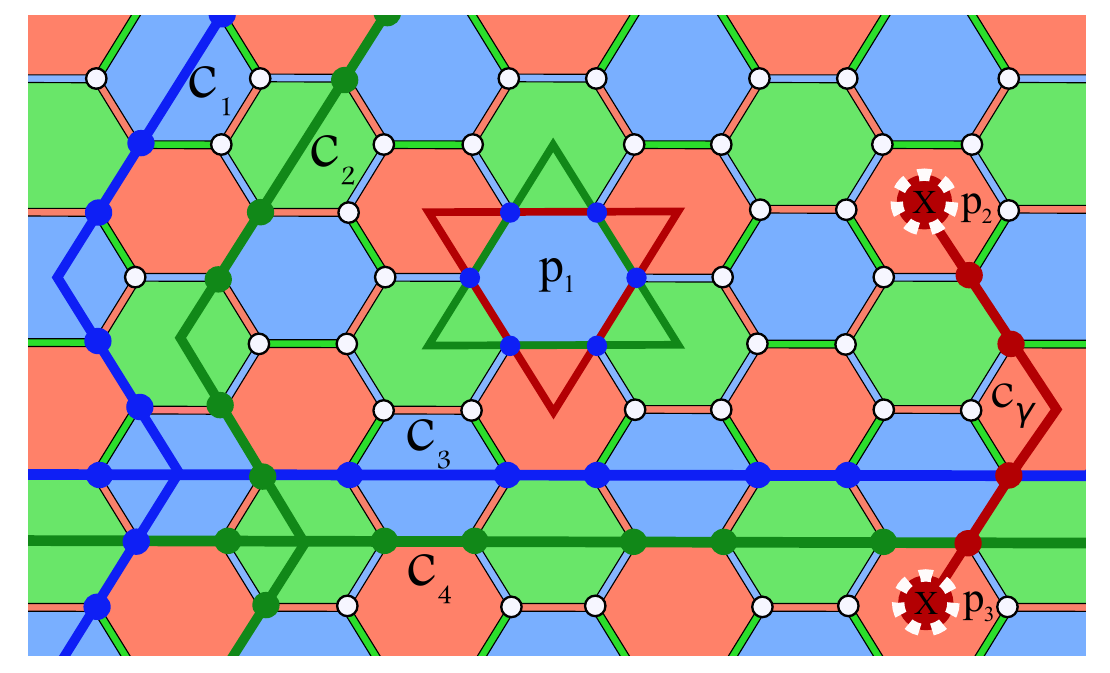}}
\caption{(Color online) TCC on the honeycomb lattice $\Lambda$ placed on a torus with genus $g=1$.
The coding subspace of the system is spanned by string operators.
A blue plaquette $p_1$ is characterized by two red and green closed strings at its boundary. An open red string (${\cal C}_{\gamma}$)
creates two $X$ particles on the surface of the two red plaquettes $p_2$ and $p_3$.
For every homology class of the torus, there are four global strings which can be labeled as (${\cal C}_1$, $\ldots$, ${\cal C}_4$).
The global strings are responsible for the topological degeneracy $16$ of the ground state.}
\label{tcc}
\end{figure}

%
%
Consider a 2D colorable lattice with a set of vertices, edges, and faces. Each vertex of the lattice is attached to three links and each link
connects two faces (plaquettes) of the same color and share the same color with the plaquettes.
Such a structure is called a 2-colex and can be visualized by three different colors
say red, green, and blue \cite{bombin1_distilation,bombin2_statistcal}. Fig.~\ref{tcc}
illustrates a piece of a 2-colex on a honeycomb lattice.
One then places spin-1/2 particles at vertices of the lattice and associate two distinct
operators $X_p=\prod_{i \in p} \sigma_i^x$ and $Z_p=\prod_{i \in p} \sigma_i^z$ to each plaquette $p$, where $\sigma_i^\alpha$'s are the usual Pauli operators.
The plaquette operators satisfy the relation $(X_p)^2=\id=(Z_p)^2$ and have eigenvalues $x_p=z_p=\pm 1$.
The Hamiltonian of the TCC reads:
\be
H_{\rm TCC}=-J\sum_{p \in \Lambda} (X_p + Z_p) \quad ,
\label{pure_tcc}
\ee
where the sum runs over all plaquettes $p$ of the lattice $\Lambda$. In the following, we set $J=1$ and we consider a honeycomb lattice.
All plaquette operators commute with the Hamiltonian (\ref{pure_tcc}) and the model is therefore exactly solvable \cite{bombin1_distilation}.
For a lattice with $N$ plaquettes, the ground-state energy per spin is $\varepsilon_0\equiv E_0/N=-2$ corresponding to $x_p=z_p=+1$ for all $p$.

Elementary excitations corresponds to $-1$ eigenvalues of the $X_p$ and $Z_p$ plaquette operators and have
the character of color. On a torus, these $X$ and $Z$ particles are created in pairs each particle increasing the energy of the system by $2$.
Consequently, the ground state is protected by this energy gap and
the system has an equidistant energy spectrum.
The elementary excitations are bosons by themselves as well as the mixture of two elementary excitations with the same color.
However, excitations of different type and color have mutual semionic statistics. In addition, combination of particles of different type
and color create two families of fermions \cite{bombin4_two_body,bombin3_anyonic_intract}.

The general form of an (unnormalized) ground state is:
\be \label{tcc_ground_state}
\ket{\Psi_{\rm{gs}}}=\prod_{p} \frac{(\id+X_p)}{2} \frac{(\id+Z_p)}{2}\ket{\upuparrows \ldots \uparrow},
\ee
where $\ket{\uparrow\uparrow\uparrow\ldots\uparrow}$ denotes a polarized spin background in the $z$-direction.
This state is a superposition of strongly fluctuating closed strings or string-net condensate \cite{levin05}.
 The ground-state degeneracy depends on the topology of the surface and for a 2-colex wrapped around a torus with $g=1$, the degeneracy
 is $16$ \cite{bombin2_statistcal,kargarian_finite_temp}.

%
%
\section{Perturbed TCC}
%
We study the model
\be \label{tcc_perturbed}
H=H_{\rm TCC}- \sum_{\alpha} \left( h_{\alpha} \sum_{i} \sigma_i^{\alpha}+j_{\alpha} \sum_{<ij>} \sigma_i^{\alpha}\sigma_j^{\alpha}\right) \quad ,
\ee
where $<ij>$ are nearest neighbor sites of $\Lambda$ and $h_\alpha>0$ ($j_{\alpha}>0$) denotes the magnetic field (Ising interaction).
We call $\alpha=x,z$ {\it parallel} and $\alpha=y$ {\it transverse} perturbation.

In the absence of perturbations, the ground state is $\Zd \times \Zd$ topologically ordered. However, for $J=0$, the ground state is disorded (polarized phase)
for a pure magnetic field while it is ordered for a pure Ising interaction. The ordered state has a $\Zd$ broken symmetry except for the isotropic
case $j_x=j_z=j$ with ($j_y\leq j$): The $SU(2)$ symmetry is broken for $j_y=j$ while the $U(1)$ symmetry of the $XY$ model is broken for $j_y<j$. Here one finds long-range order and gapless excitations.
Consequently, a phase transition out of the $\Zd \times \Zd$ topologically ordered phase has to occur when either a magnetic field or an Ising interaction is
sufficiently strong.

\section{Methods}

\subsection{Perturbative continuous unitary transformations}
Our aim is to study the low-energy spectrum of the TCC and dynamics of the model in the presence of perturbations. To this end, 
we use the perturbative continuous unitary transformations method which was developed by Knetter and Uhrig \cite{pcut1,pcut2} for many-body Hamiltonians.
The pCUT method computes the perturbative expansion of the Hamiltonians which can be written in the form
\be
H=Q+\sum_{n=-N_{max}}^{N_{max}} T_n
\ee
provided that: {\bf i}) The unperturbed part of the Hamiltonian, $Q$, has a simple and equidistant spectrum bounded from below and  {\bf ii}) the perturbed part
of the Hamiltonian can be recast in terms of $T_n$ operators where increment 
(decrement, if $n < 0$) the number of excitations (quasiparticles) by $n$ such that $\comutator{Q}{T_n}=nT_n$.
The pCUT method then maps the initial Hamiltonian to an effective one which conserves the number of quasiparticles and is expressed as follows:
\be
H_{\rm eff}=Q+\sum_{k=1}^{\infty}x^{k} \sum_{\overline{m}=0 }C(m_1 \ldots m_k) T_{m_1}\ldots T_{m_k} \quad,
\label{effectiv_h}
\ee
where the first sum runs over the order of perturbation, $k$, 
and the second one runs over all possible permutations of $\{m_1,m_2,\ldots,m_k\}$ with $m_i \in \{-N_{max}, \ldots, N_{max}\}$ which satisfy the 
condition $\overline{m}=\sum_i m_i=0$. 
The coefficients $C(m_1 \ldots m_k)$ are computed as exact rational numbers in different orders of perturbation\cite{pcut2}.
The $T_{m}$ operators in order $k$ are responsible for the dispersions and fluctuations of the particles on the lattice. 
Note that the pCUT method provides the results which are exact up to the computed order of perturbation in the thermodynamic limits.

\begin{figure}
\centerline{\includegraphics[width=\columnwidth]{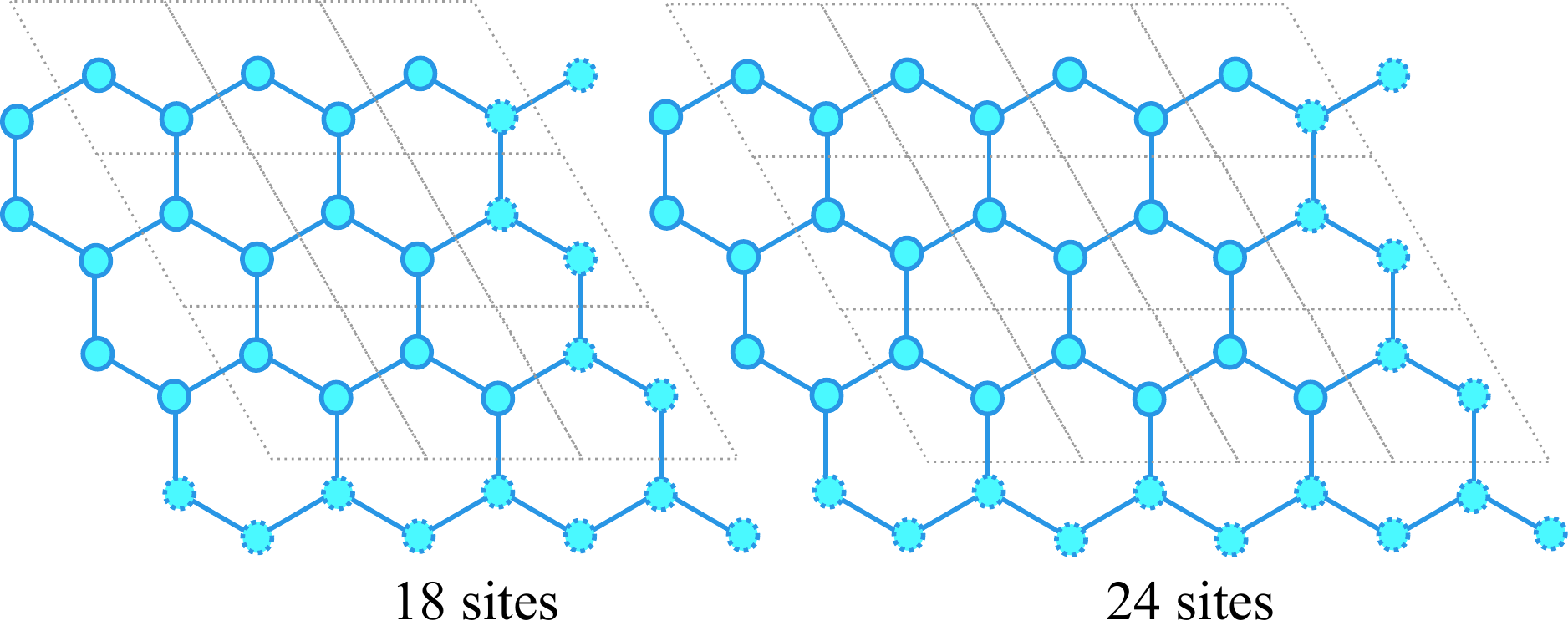}}
\caption{(Color online) Honeycomb clusters with periodic boundary condition. Closed circles are the physical spins at the vertices of the lattice and the dashed 
circles denote the periodicity at the boundaries of the lattice. The dotted gray lines further demonstrate the two site unit cells which span the lattice.}
\label{cluster_honey}
\end{figure}

\subsection{Exact diagonalization}

As the second approach, we use the finite-size exact diagonalization method to calculate the energy spectrum of the system.
We use the $z$-Pauli spin vector space and bit representation to generate the matrix elements of the Hamiltonian and extract the low-lying energy spectrum,
using the Lanczos algorithm. The numerical diagonalization is performed on $L\times M$ honeycomb clusters with $L$ ($M$) being the number of two-site unit cells along 
the height (width) of the honeycomb clusters (see Fig.~\ref{cluster_honey}). 
Imposing periodic boundary conditions at the boundaries of the clusters, we performed the ED on a $3\times3$ (18 sites) and a $3\times4$ (24 sites) clusters.
\begin{figure}
\centerline{\includegraphics[width=4cm]{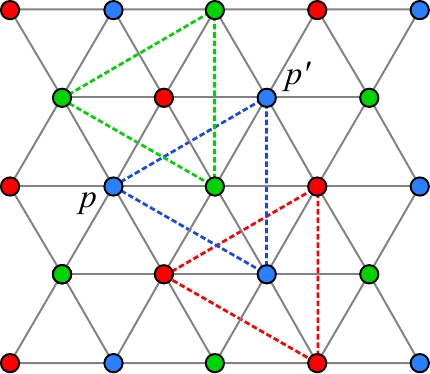}}
\caption{(Color online) TCC on the dual triangular lattice. The colored vertices represent the plaquettes of the same color on the original honeycomb lattice.
A single parallel Ising perturbation is mapped to three Ising interaction on the nearest-neighbor sites 
$p, p'$ of the colored triangular sublattices formed by connecting the vertices of the same color.}
\label{dual_latt}
\end{figure}
%
\section{Parallel perturbations}
%
%
\begin{figure}
\centerline{\includegraphics[width=\columnwidth, height=6cm]{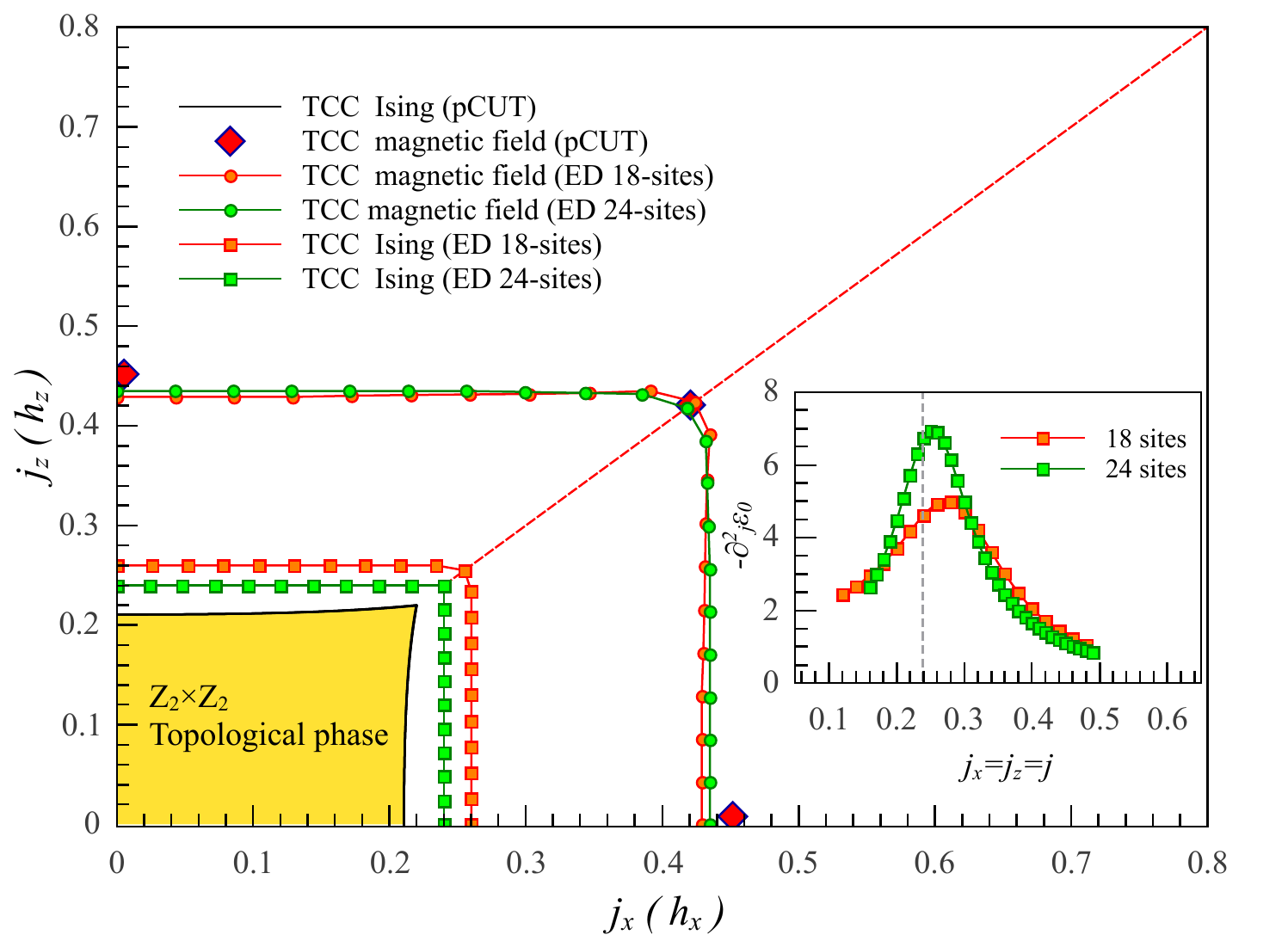}}
\caption{(Color online) Phase diagram of the TCC perturbed by parallel perturbations $(j_x,j_z)$ or $(h_x,h_z)$ setting $J=1$.
The boundaries of the yellow region correspond to second-order phase transitions of the TCC in the presence of Ising interactions $(j_x,j_z)$ obtained from the 1-QP gap 
in the small-coupling limit. The circles are the corresponding phase boundaries obtained by ED on a 18- and 24-sites periodic clusters.
The red dashed line and squares represent first-order phase transitions calculated by ED on periodic 18 and 24-site clusters. 
The red diamonds on the axis and on the isotropic line ($h_x=h_z$) are the corresponding first-order points obtained by pCUT \cite{ssj}.
The inset further shows how the increase in system size sharpens the resonances of the $-\partial^2_j\varepsilon_0$ for $j_x=j_z=j$ in ED calculations.}
\label{phase_diagram}
\end{figure}

We start with the simplest case of a single parallel perturbation $h_x$ or $j_x$ (the case $h_z$ or $j_z$ is identical up to an interchange of $X$ and $Z$
particles). The $X_p$ plaquette operators remain conserved quantities, i.e.~the Hilbert space decouples into subspaces having
fixed configurations of $x_p$ eigenvalues. The low-energy physics is always contained in the sector where $x_p=+1$ for all $p$ \cite{ssj}.

The TCC in a single field $h_x$ can be mapped to the Baxter-Wu model \cite{baxter_wu} in transverse
magnetic field by a duality transformation. The first-order phase transition between the topologically ordered phase and the polarized phase is
located at $h_x^{\rm c}\approx 0.383$ \cite{ssj}.

\begin{figure}
\includegraphics[width=\columnwidth, height=5.5cm]{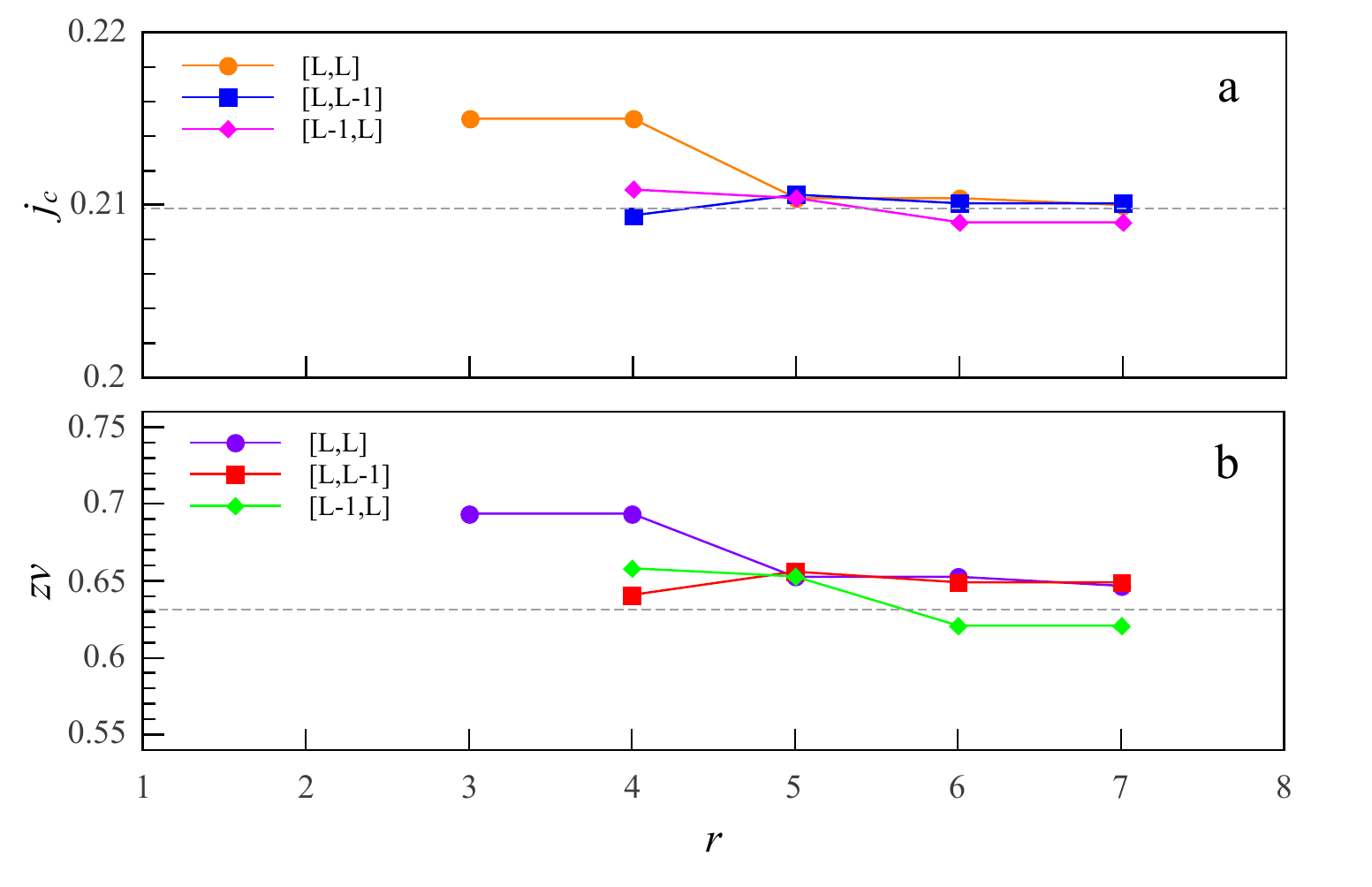}
\caption{(Color online) Transition point $j_c$ (a) and the critical exponent $z\nu$ (b) of the triangular TFIM extrapolated obtained from DlogPad\'{e} approximants as a function of the order $r$ of the series for $J=1$. Here $L$ is given by $L = $[$r-(r $mod$ 2)$]$/2$. 
The literature values $j_c = 0.209$ and $z\nu=0.630$ \cite{He,bloete95} are highlighted by dashed lines.}\label{ising_limit_critical}
\end{figure}

An analog duality mapping exists for the TCC plus $j_x$ Ising interaction. One introduces pseudo-spin 1/2 operators $\tau_p^\alpha$ with $\alpha=x,y,z$ living on the 
dual triangular lattice of plaquettes. Apart from an unimportant constant due to the $x_p$ eigenvalues, the energetics of the TCC is mapped to an effective field term,
\be
H_{\rm TCC}^{\rm dual}=-NJ-J\sum_p \tau^z_p
\ee
where the sum runs over the vertices of the dual triangular lattice.
A single $\sigma_i^x$ flips the $z_p$ eigenvalue of all three plaquettes $p$ which contain the site $i$ \cite{ssj}. The Ising interaction 
$j_x$ then gives rise to an effective Ising interaction,
\be
H_{\rm TCC}^{\rm map}=-NJ-J\sum_p \tau^z_p-j_x\sum_{c}\sum_{\langle p,p'\rangle_c} \tau^x_p\tau^x_{p'}
\ee
where the inner sum runs over the nearest-neighbor vertices (plaquettes) of the same color $c$ being red, green, and blue on the triangular (honeycomb) 
lattice (see Fig.~\ref{dual_latt}). One obtains three independent TFIMs on the dual triangular lattice
of plaquettes with the same color. The triangular TFIM displays a second-order phase transition. Consequently, we can establish rigorously a quantum phase transition in the 3D Ising universality class between a $\Zd \times \Zd$ topologically ordered phase and a $\Zd$ symmetry-broken phase which takes place 
at \mbox{$j_{x}^{\rm c}\approx 0.209$} \cite{He}, i.e. all critical exponents correspond to an 3D Ising transition. Physically, the quantum phase transition corresponds to a condensation of $Z$ particles which gain kinetic energy due to the presence of $j_x$. Using high-order series expansions, we calculated the 1-QP gap of the system and we recovered the results of Ref.~\onlinecite{He}. 

The TCC in a general parallel magnetic field ($h_x$,$h_z$) yields a rectangular shape of the topological phase boundary 
(see Fig.~\ref{phase_diagram}). The transition is first order for all field directions which is detected by a jump of $\partial_h\varepsilon_0$. 
We used ED for a general field direction $(h_x,h_z$) which shows only little finite-size effects for a first-order phase transition. Additionally, we have performed pCUT about both limits for the case $h_x=h_z$ giving a first-order phase transition at $h_x=h_z\approx 0.42$. The remaining (small) differences between pCUT and ED data on the axis as well as for $h_x=h_z$ are due to finite-size effects which become smaller for increasing system size.

The physics is different for the case ($j_x$,$j_z$). We calculated $\varepsilon_0$ as well as
the 1-QP dispersion $\omega(\bf{k})$ and the 1-QP gap \mbox{$\Delta=\omega(\bf{k}=0)$} using pCUT. The gap is given by
the $Z$ particle ($X$ particle) for \mbox{$j_z<j_x$} ($j_z>j_x$). Both gaps are degenerate for $j_x=j_z$. The gap behaves
as $\Delta\sim(j-j^{\rm c})^{z\nu}$ in the vicinity of the quantum critical point $j^{\rm c}$ where $z$ is the dynamical critical exponent and 
$\nu$ is the correlation length exponent. Setting $j_x=0$ (or $j_z=0$), we recover the results obtained in Ref.~\onlinecite{He} for the triangular TFIM. 
Using DlogPad\'e approximants, we obtain $z\nu\approx 0.65$ which is fully consistent with the expected 3D Ising exponents $\nu\approx 0.63$ \cite{bloete95} and $z=1$ (see Fig.~\ref{ising_limit_critical}).

\begin{figure}
\includegraphics[width=8.5cm,height=5.5cm]{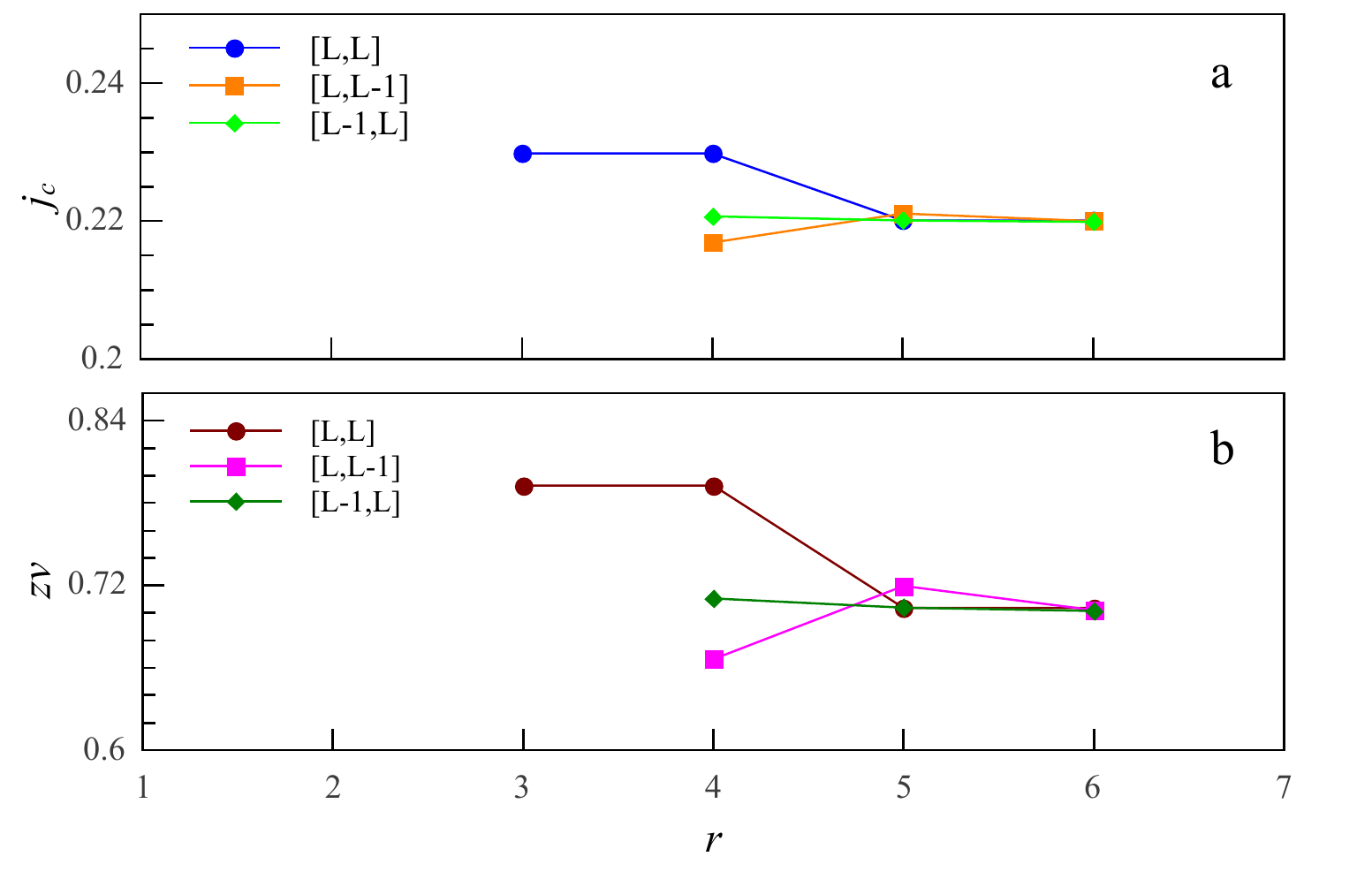}
\caption{(Color online) Transition point $j_c$ ($J=1$) and critical exponent $z\nu$ of the TCC at the multicritical point \mbox{($j_x=j_z,j_y=0$)} extracted 
from the DlogPad\'{e} approximants [$L,L$], [$L+1,L$] and [$L,L-1$] with $L = $[$r-(r $mod$ 2)$]$/2$ of the one-particle gap $\Delta$ obtained by pCUT.} \label{hx_hz_critical} 
\end{figure}

When both Ising interactions $j_x$ and $j_z$ are finite, the analysis of the 1-QP gap predicts two critical lines which connect to the Ising critical points on the $j_x$ ($j_z$) axis and intersect at the point $j_x=j_z\approx 0.22$. Along the critical lines, the critical
exponent $z\nu$ stays close to the 3D Ising exponent except at the multicritical point where one finds $z\nu\approx0.7$ (see Fig~\ref{hx_hz_critical}).
 Keeping in mind the exact duality to the TFIM on the triangular lattice for $j_x=0$ ($j_z=0$), one therefore expects a 3D Ising universality class for all cases except $j_x=j_z$. Here the different critical behavior is likely a consequence of the fact that the semionic $X$ and $Z$ particles
condense simultaneously for $j_x=j_z$. Interestingly, a very similar criticality is found for the toric code in a parallel magnetic field \cite{toric_dusuel}. 

Our findings for the parallel Ising interactions $j_x$ and $j_z$ are also consistent with our ED data. As expected, finite-size effects are larger compared to the first-order phase transitions discussed above for parallel magnetic fields. Furthermore, the second derivative of the ground-state energy $\varepsilon_0$ displays a characteristic resonance, which sharpens with increasing system size, for Ising interactions consistent with the critical lines deduced by series expansion (see inset in Fig.~\ref{phase_diagram}).

Outside the topological phase, the ground state of the system is $\Zd$ symmetry-broken except for $j_x=j_z$ where the perturbation
is the $XY$ model. Our ED results unveil that the system undergoes a first-order transition between both $\Zd$ symmetry-broken phases
when crossing the line $j_x=j_z$ up to the limit $J=0$.

\section{Transverse perturbations}
For a pure transverse perturbation, $X_p$ and $Z_p$ plaquette operators are no longer conserved quantities.
The operator $\sigma_i^y$ anticommutes with both $X_p$ and $Z_p$ on the three plaquettes containing the site $i$ and flips the eigenvalues of the plaquette operators, simultaneously. 

\begin{figure}
\centerline{\includegraphics[width=\columnwidth]{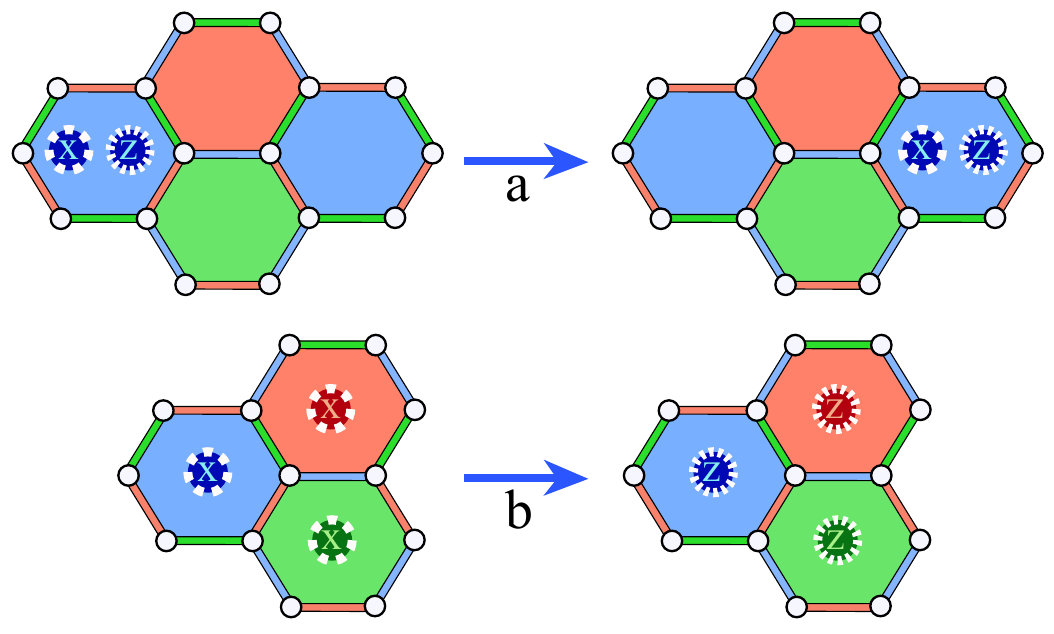}}
\caption{(Color online) (a) A two-particle bound state with even particle number parity. The two bounded particles hop to the neighboring blue plaquette in order two perturbation theory.
(b) A three-particle bound state with odd parity on the plaquettes. The three bounded $X$-type QPs turn together into $Z$-type particles (or vice verse) in odd orders of perturbation.}
\label{bound_state}
\end{figure}

The pCUT maps the model to a quasiparticle conserving effective Hamiltonian, allowing to investigate the low-energy properties of the system and the dynamics of QPs.
The effective Hamiltonian not only conserves the number of QPs, but, more interestingly, $H_{\rm eff}$ also conserves the parity of the number of
quasiparticles on the surface of the plaquettes. 
The combined effect of quasiparticle and parity conservation has important consequences on the properties of QPs inside the topological phase, 
e.g.~a single $Z$ or $X$ QP is strictly local since all plaquettes except one have an even (zero) number of QPs and therefore any hopping of the 
QP violates the parity conservation.

Furthermore, one finds bound states due to the attractive interaction between $X$ and $Z$ particles induced by the transverse perturbations.
Indeed, the elementary excitation having a true two-dimensional dispersion is the composite
2QP object having one $X$ and one $Z$ particle on the same plaquette. The parity conservation forces the two QPs to stay attached to each
other and move together on the dual triangular lattice of plaquettes having the same color. An example of a two bounded quasiparticles is illustrated in Fig.~\ref{bound_state}-(a).
In the pCUT language, a quasi-particle conserving process such as $T_{-2}T_{2}$ in order two perturbation theory is responsible for the hopping of bounded QPs from the 
left blue plaquette to the right blue one.

In a similar manner, three bounded particles on the surface of the three neighboring plaquettes such as the one shown in Fig.~\ref{bound_state}-(b) stays local due to the 
odd parity of the plaquettes. However, the action of a $T_0$ process corresponding to first order perturbation theory can transform the $X$ and $Z$ QPs to each other.

\begin{figure}
\centerline{\includegraphics[width=\columnwidth, height=6cm]{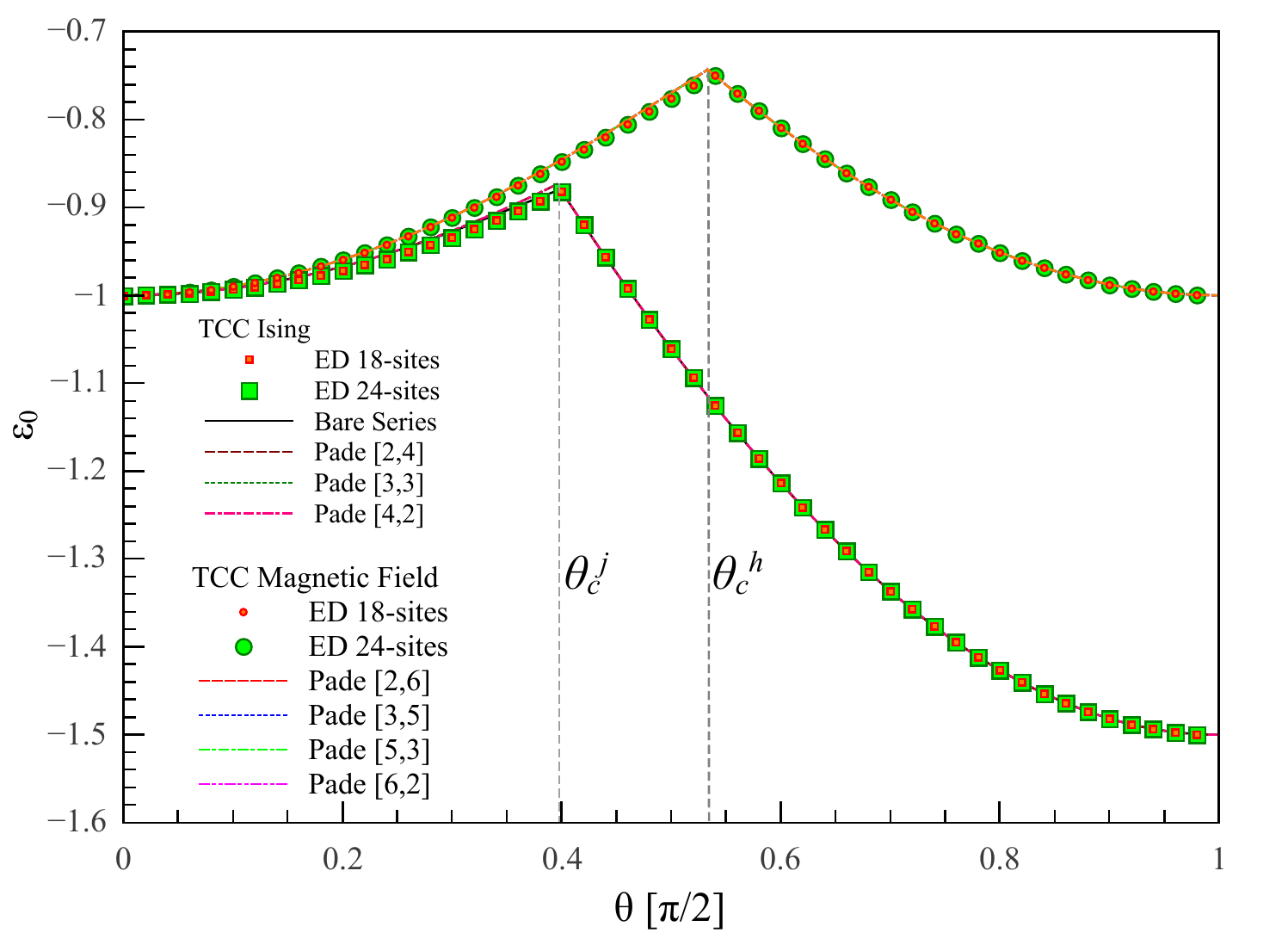}}
\caption{(Color online) The ground-state energy per spin $\varepsilon_0$ of the TCC plus transverse perturbations $j_y$ or $h_y$ as a
function of $\theta$ in units of $\pi/2$ setting $J=\cos\theta$ and $j_y,h_y=\sin\theta$. The discontinuities in $\varepsilon_0$ signal first-order phase transitions at \mbox{$\theta^{j}_c\approx 0.62$} for $j_y$ and at \mbox{$\theta^{h}_c\approx 0.84$} for $h_y$ (indicated by vertical dashed lines).}
\label{gsps_y_yy}
\end{figure}

In the following, we continue our discussion by focusing on the high-field (large-coupling) limit of the problem.
The ground state of the system in $J=0$ limit corresponds to fully polarized states
in $\pm y$-direction. Consequently, one can look at the limit $J\neq0$ by setting up a perturbative picture.
Noting that $\sigma^z=i\sigma^y\sigma^x$, the Hamiltonian of TCC in transverse perturbation can be recast into:
\be \label{transverse_tcc_hf}
H=-h_y \sum_i \sigma_i^y-j_y \sum_{<ij>} \sigma_i^y\sigma_j^y-J\sum_p X_p(1 - Y_p) \quad ,
\ee
As we can clearly see from the right term of the above Hamiltonian, the polarized ground state of the system in $J=0$ limit is
the simultaneous eigenstate of the $Y_p$ plaquette operator with eigenvalue $+1$ and as a result the action of
the pure TCC on the polarized ground states is zero. More precisely, the 0-QP state is an exact eigenstate
of the full Hamiltonian (\ref{transverse_tcc_hf}) and the Hamiltonian is diagonal in this eigenbasis. The ground-state
energy of the system in the large-coupling phase is therefore $-Nh_y$ ($-(3/2) N j_y$) where $N$ is the number of sites on the honeycomb lattice.

We computed $\varepsilon_0$ for the small-coupling limit inside the topological phase for both transverse
perturbations $h_y$ and $j_y$ using pCUT. Additionally, we have performed ED on a periodic honeycomb cluster with 18 and 24-sites.
Our results reveal first-order phase transitions for both transverse perturbations taking place
at $h_y^{\rm c}\approx 0.74$ and \mbox{$j_y^{\rm c}\approx 0.58$} (see Fig.~\ref{gsps_y_yy}). Physically, this is reasonable keeping in mind that the exactly known ground state at large couplings contains no quantum fluctuations.

\begin{figure}
\includegraphics[width=\columnwidth, height=6cm]{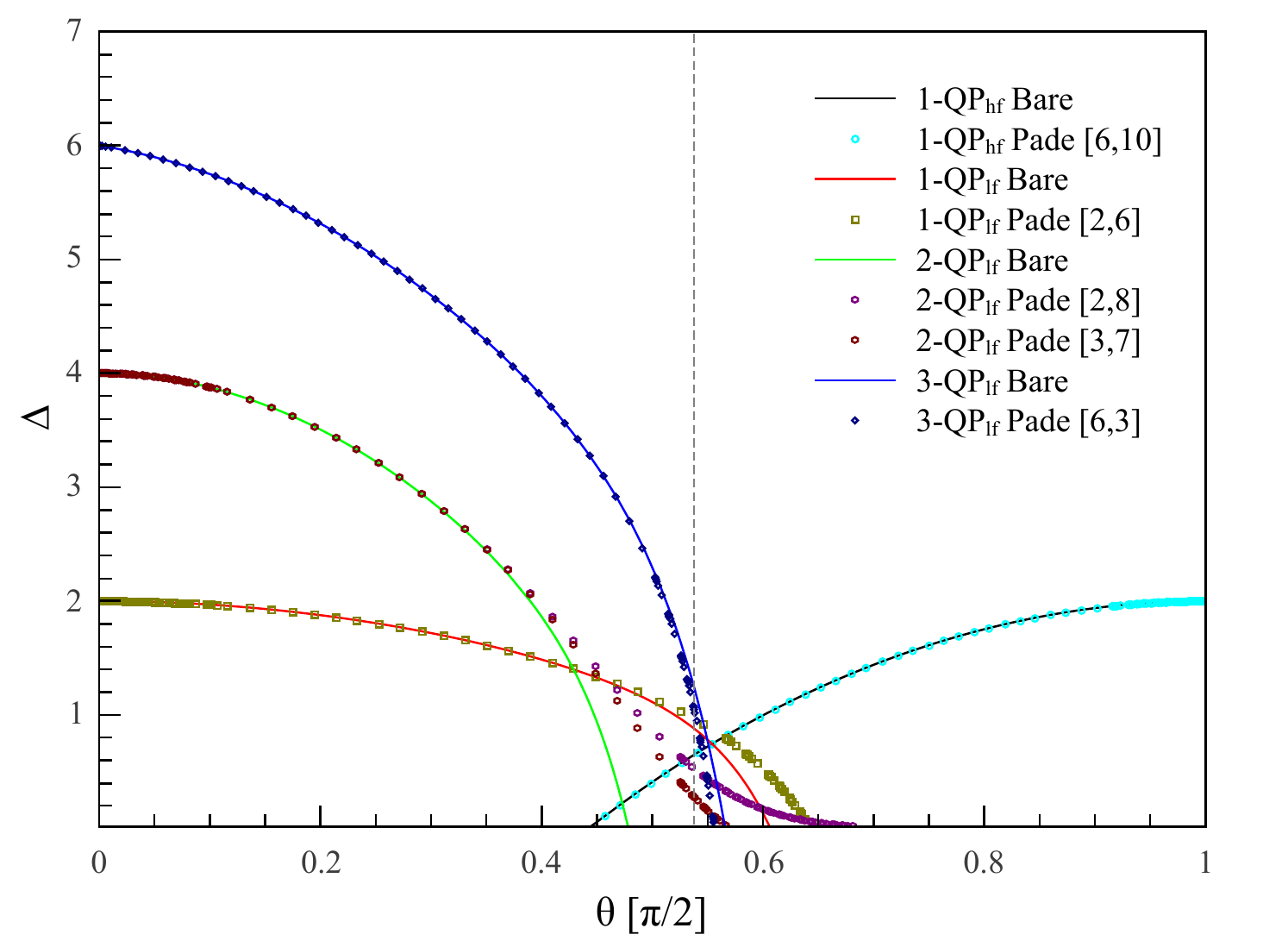}
\caption{(Color online) Low- and high-field gap $\Delta$ of the TCC in transverse magnetic field as a function of $\theta$ such that $J=\cos\theta$ and 
$h_y=\sin\theta$. Solid lines correspond to the bare series of maximum order 8, 10, and 9 for 1-, 2- and 3-QP low-field gaps, respectively and 
order 16 for high-field 1-QP gap. The symbols also denote different Pad\'{e} approximants. The vertical gray dashed line represents the transition point 
obtained by analyzing ground-state energies (see Fig.~\ref{gsps_y_yy}).}
\label{gap_y}
\end{figure}

We have further computed the dispersion and the 1-, 2- and 3-QP gaps of the system up to high orders in the expansion parameter $h_y$.
Let us stress that the 2-QP and 3-QP gaps are computed for the situations identical to those of Figs.~\ref{bound_state}-(a) and (b), respectively.
Interestingly, the results from the low- and high-field gaps are also consistent with first-order transition (see Fig.~\ref{gap_y}). The 1- and 3-QP low-field gaps
are well converged and their intersection with high-field gap occurs at a value very close to that of the intersection of the ground-state energy per site,
while due to its alternating signs, the 2-QP low-field gap has a relatively poor convergence and its crossing point deviates from $\theta_c$.

\section{Isotropic plane ($j_x=j_z$)}

Let us merge our results for the parallel and transverse Ising perturbation by
 considerung the parameter plane ($j,j_y$) with $j\equiv j_x=j_z$. This plane is of particular
 interest, since i) excitation energies of $X$ and $Z$ particles are exactly degenerate and ii)
 the model exhibits gapless phases for $j\leq j_y$ in the large-coupling limit $J=0$.

\begin{figure}
\centerline{\includegraphics[width=\columnwidth, height=6cm]{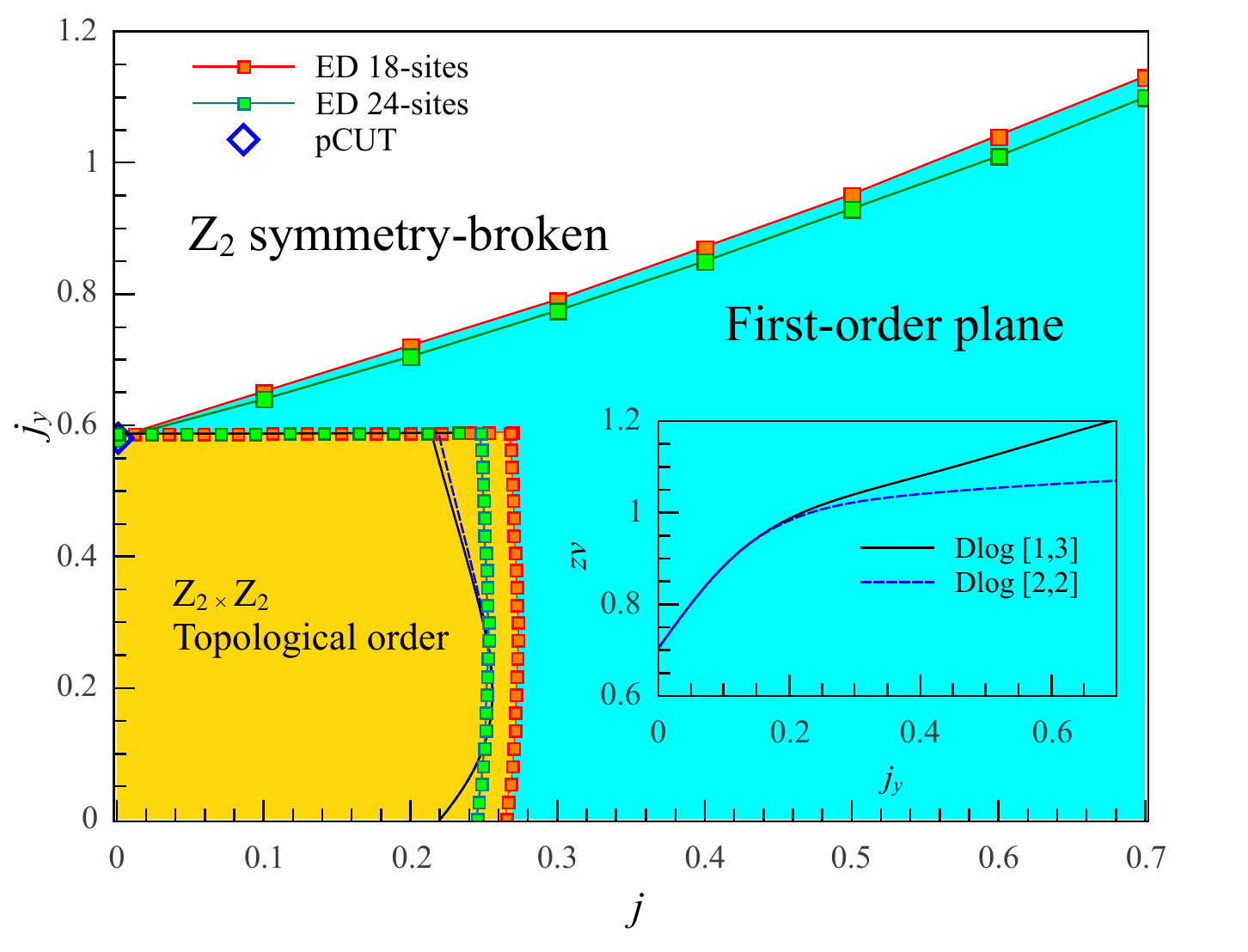}}
\caption{(Color online) Phase diagram of the TCC plus Ising interactions ($j,j_y$) with $j\equiv j_x=j_z$ and $J=1$. Solid lines are obtained by pCUTs while red and green
symbols correspond to ED data of periodic 18- and 24-site clusters, respectively. 
The blue diamond represents the transition point obtained by pCUT for ($0,j_y$).
The yellow region illustrates the topologically ordered phase while the cyan/gray shaded area represents a plane of first-order phase transitions. 
{\it Inset}: Critical exponent $z\nu$ as a functions of $j_y$ obtained from different DlogPad\'{e} extrapolants $[L,M]$ of the 1-QP gap.}
\label{isotropic-plane}
\end{figure}

Combining pCUT results in the topological phase and ED, we obtain the
 phase diagram shown in Fig.~\ref{isotropic-plane}. Let us stress that, although critical properties
 are clearly influenced by finite-size effects, we expect that first-order
 phase transitions are almost converged on the finite cluster 
( as already found for the parallel and transverse perturbations discussed above).

The topologically ordered phase is bounded by one first-order line and one second-order line. The first-order line is almost straight starting at $(0,j_y^{\rm c})$  and 
ending at \mbox{$\approx(0.22,j_y^{\rm c})$} where it meets the second-order line which is adiabatically connected to the multicritical point 
\mbox{$\approx(0.22,0$)}. Interestingly, DlogPad\'{e} extrapolants of the 1-QP gap indicate the critical exponent $z\nu$ to vary continuously between 
$0.7$ and $\approx 1$ on this multi-critical line which is in striking similarity to the phase diagram of the toric code in a field \cite{toric_dusuel}. 
Although the convergence is still poor for this perturbative order and finite-size effects in ED are still significant, our results suggest that the semionic statistics is essential to understand the quantum criticality 
in both models.

In contrast, the physics is clearly different to the perturbed toric code outside the topological phase. Here we find a plane of first-order transitions, 
i.e.~when going at a finite angle through any point of the grey shaded area displayed in Fig.~\ref{isotropic-plane} one finds a first-order transition. 
The  upper end of this first-order plane is then a critical line separating the gapped $\Zd$ broken phase at larger Ising interaction $j_y$ from the first-order plane.
We expect that this end line approaches the isotropic line $j=j_y$ in the limit of infinite Ising interactions. 
Let us recall that for $J=0$, one has a second-order transition at the $SU(2)$ symmetric Heisenberg point $j=j_y$ when approaching from  
the $\Zd$ broken phase at $j<j_y$. Finally, we point out that the ED detects no further phase transitions inside the grey-shaded region. 
The latter suggests that the multicritical line separates the topologically ordered phase at small couplings from a long-range ordered gapless phase adiabatically 
connected to the large-coupling limit which is certainly a very interestering scenario.

\begin{figure}
\includegraphics[width=\columnwidth]{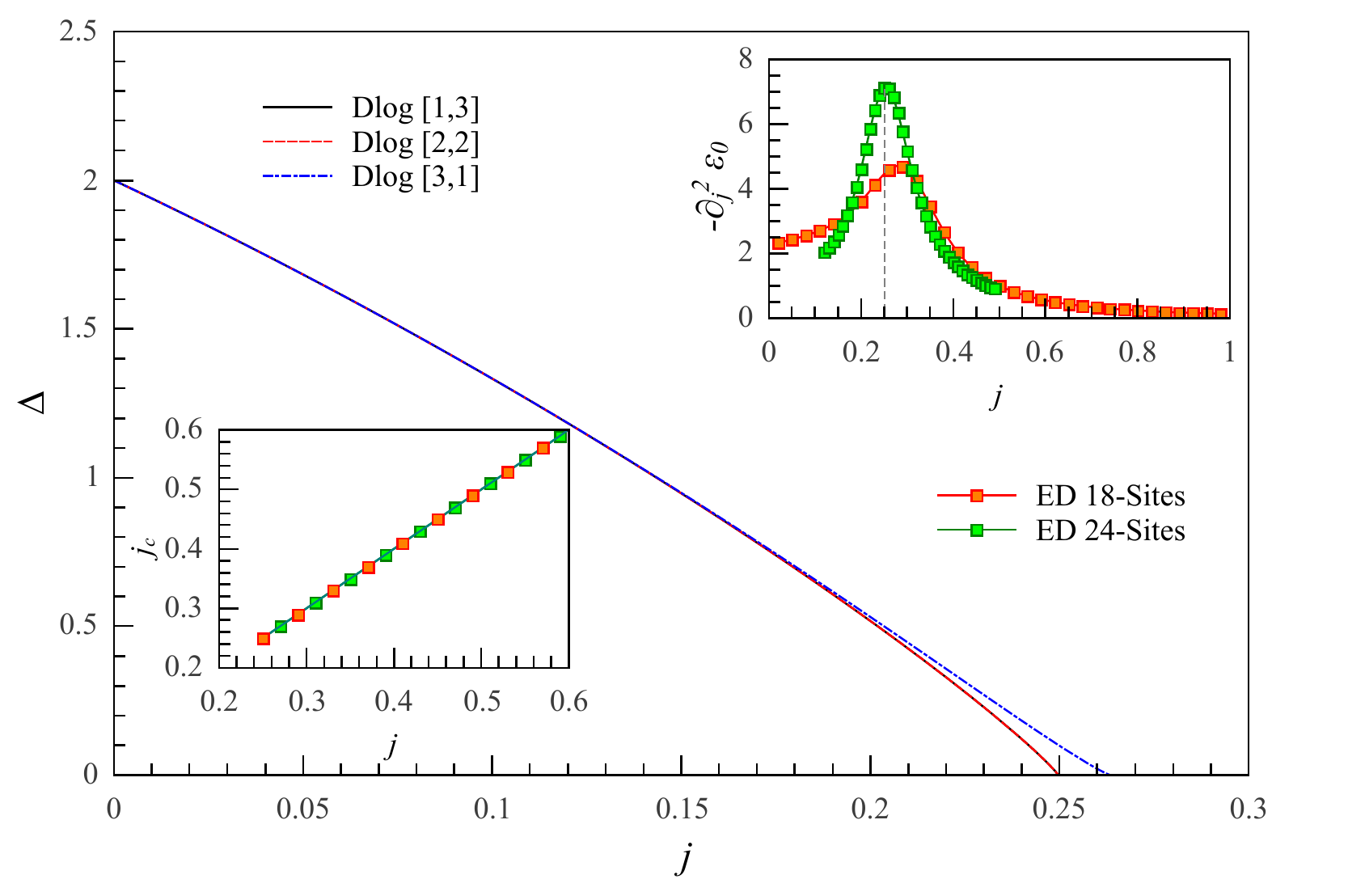}
\caption{(Color online) The 1-QP gap of the perturbed TCC on the isotropic line ($j\equiv j_x=j_y=j_z$) for different DlogPad\'{e} approximants. Here we have set $J=1$.
The upper inset depicts $-\partial^2_j\varepsilon_0$ obtained from ED for periodic clusters with 18 and 24 sites. 
The critical point highlighted by gray dashed line is consistent with the closure of the gap at $j_c\approx 0.25$. 
The lower inset displays the first-order transition points on the isotropic line which merge into the second-order point obtained by the closure of the gap.}\label{gap_iso}
\end{figure}

In the case where all of the Ising interactions are equal ($j\equiv j_x=j_y=j_z$), the perturbation corresponds to the isotropic Heisenberg model. 
At very large couplings ($j\gg 0$), the ground state of the system has a long-range ferromagnetic order and the low-lying excitations are
gapless magnons with broken $SU(2)$ symmetry. The system therefore undergoes a quantum phase transition between the topological phase and the symmetry-broken phase.
Closure of the 1-QP gap from the small-coupling limit of the problem signals a second-order transition at $j_c\approx0.25$ (see Fig.~\ref{gap_iso}). 
Additionally, the ED analysis on honeycomb periodic clusters with 18 and 24 sites further reveals that the ground state susceptibility, derived by the Hellman-Feynman 
theorem: ($\chi=-\partial^2_j \varepsilon_0$), gives rise to a resonance at $j_c\approx0.26$
which is in full agreement with the closure of the gap and a second-order phase transition (see upper inset of Fig.~\ref{gap_iso}). 
This second-order point further merges into a first-order line which lies inside the isotropic plane and continues up to the limit $J=0$ 
(see lower inset of Fig.~\ref{gap_iso}).

\section{Conclusions}

We studied topological phase transitions of the TCC in the presence of a uniform magnetic field or Ising interactions on a honeycomb lattice.
For the TCC in a uniform magnetic field we find first-order transitions to a polarized phase for all field directions. 
In the presence of a single parallel field, the system is mapped to the Baxter-Wu model in transverse field and undergoes a first-order transition at $h_x^c\approx0.383$
\cite{ssj}. When the two parallel fields are turned on, the phase diagram of the model has a rectangular shape in which the phase boundaries originate from the 
Baxter-Wu limit on the axis and merge into a transition point at $h_x=h_z\approx0.42$ on the isotropic line.

In contrast, the TCC in single parallel Ising interaction is mapped to the transverse field Ising model on the triangular lattice which undergoes a second-order phase transition
between the $\Zd \times \Zd$ topologically ordered phase of the TCC and conventional symmetry-broken phase of the Ising model at $j_x\approx0.209$ in the 3D Ising universality class.
When the two parallel Ising interactions are nonvanishing, the phase boundaries are two second-order lines which start from the Ising limit and merge into a multicritical point at 
$j_x=j_z=0.22$. The $z\nu$ critical exponent of the system stays close to the Ising value and starts to grow near the multicritical point until it reaches $\approx0.7$
at $j_x=j_z$. Outside the topological phase, the ground state of the system is $\Zd$ symmetry-broken except for $j_x=j_z$ where the perturbation is the $XY$ model.
This limit of the problem becomes very interesting when we turn on the transverse perturbation. 
The first order line outside the topological phase becomes a first order plane and the critical exponent $z\nu$ starts to vary continuously between 
$0.7$ and $\approx 1$ which is similar to the behavior of the toric code in a field \cite{toric_dusuel}.

The topological phase of the TCC breaks down to a polarized phase in pure traverse magnetic field and Ising perturbation at $h_y^c\approx0.74$ and $j_y^c\approx0.58$.
In this limit, neither of the $X$ and $Z$ plaquette operators are conserved quantities. Consequently the system exhibit interesting characteristics such as the conservation of the 
parity of the number of particles on the plaquettes and formation of bounded quasiparticles.
Moreover, the polarized phase is the exact eigenstate of the perturbed TCC in transverse perturbations.

\section{Acknowledgements}
K.P.S. acknowledges ESF and EuroHorcs for funding through his EURYI. M.K. acknowledges financial support from ARO Grant NO. W911NF-09-1-0527 and NSF Grant NO. DMR- 0955778.
S.S.J. also acknowledges KPS for hospitality during his stay at the TU Dortmund.

\onecolumngrid
\newpage
\appendix

\section{Series expansion of the perturbed TCC with Ising interaction}
Here we present the series expansions of ground-state energy per site, $\varepsilon_0$, and the one-particle gap, $\Delta=\omega(\bf{k}=0)$,
of the system obtained from pCUT for the small-coupling limit of the perturbed TCC with Ising interaction. The gap $\Delta$ given below is for one $Z$ quasi particle
which is the true gap of the system for $j_x\geq j_z$. For $j_z> j_x$, the true gap is the one of $X$ quasi particles which can be obtained from $\Delta$ by interchanging $j_x$ and $j_z$. Note that we have set $J=1/2$ for both series:

\bea \label{gs-ising-small_paral}
\varepsilon_{0} &=& -\frac{1}{2}-\frac{3}{8} (j_x^2+j_z^2)-\frac{3}{16}j_y^2-\frac{3}{4} (j_x^3+j_z^3-j_x j_z j_y)+\frac{3}{16} j_y^3-\frac{87}{32} (j_x^4+j_z^4)-\frac{87}{256} j_y^4 \nonumber\\
&& -\frac{31}{128} (j_x^2 j_y^2+j_z^2 j_y^2)+\frac{9}{4} (j_x j_z^2 j_y+j_x^2 j_z j_y)+\frac{3}{4} j_x^2 j_z^2-\frac{39}{32} j_x j_z j_y^2-\frac{99}{8} (j_x^5+j_z^5)+\frac{99}{128} j_y^5 \nonumber\\
&& +\frac{57}{16} (j_x^3 j_z^2+j_x^2 j_z^3)+\frac{265}{768} (j_x^2 j_y^3+j_z^2 j_y^3)+\frac{75}{8} (j_x^3 j_z j_y+j_x j_z^3 j_y)-\frac{877}{768} (j_x^3 j_y^2+j_z^3 j_y^2)  \nonumber\\
&& -\frac{621}{128} (j_x j_z^2 j_y^2+j_x^2 j_z j_y^2)+\frac{27}{4} j_x^2 j_z^2 j_y+\frac{2477}{768} j_x j_z j_y^3  \quad. \nonumber
\eea

\bea \label{gap-ising-small_paral}
\Delta &=& 1-6 j_z+12 j_x^2 j_z+24 j_x^3 j_z+168 j_x^4 j_z-12 j_z^2+33 j_x^2 j_z^2+\frac{363}{4} j_x^3 j_z^2-42 j_z^3+\frac{945}{4} j_x^2 j_z^3-252 j_z^4 \nonumber\\
&& -\frac{3153}{2} j_z^5+18 j_x^2 j_y+75 j_x^3 j_y+435 j_x^4 j_y+24 j_x j_z j_y+72 j_x^2 j_z j_y+282 j_x^3 j_z j_y+90 j_x j_z^2 j_y \nonumber\\
&& +\frac{441}{2} j_x^2 j_z^2 j_y+765 j_x j_z^3 j_y-\frac{3}{2} j_y^2-\frac{51}{4} j_x j_y^2-\frac{2539}{32} j_x^2 j_y^2-\frac{40873}{96} j_x^3 j_y^2+\frac{39}{8} j_z j_y^2-\frac{465}{8} j_x j_z j_y^2 \nonumber\\
&& -\frac{97603}{192} j_x^2 j_z j_y^2-\frac{147}{8} j_z^2 j_y^2-\frac{10027}{48} j_x j_z^2 j_y^2-\frac{8275}{384} j_z^3 j_y^2+\frac{15}{4} j_y^3+\frac{273}{8} j_x j_y^3 \nonumber\\
&& +\frac{25243}{96} j_x^2 j_y^3-\frac{39}{4} j_z j_y^3+\frac{13951}{64} j_x j_z j_y^3+\frac{4665}{64} j_z^2 j_y^3-\frac{879}{8} j_x j_y^4+\frac{2505}{64} j_z j_y^4+\frac{159}{4} j_y^5 \quad. \nonumber
\eea

\section{Series expansion of the TCC in parallel magnetic field}

Ground-state energy per site $\varepsilon^{\rm lf}_{0}$ and 1-QP gap $\Delta^{\rm lf}$ of the TCC in parallel magnetic field in the low-field limit for $J=1/2$:
\bea \label{gs-small_paral}
\varepsilon^{\rm lf}_{0} &=& -\frac{1}{2}-\frac{1}{6}(h_x^2+h_z^2)-\frac{19}{108}(h_x^4+h_z^4)+\frac{1}{27} h_x^2 h_z^2 -\frac{10718}{8505}(h_x^6+h_z^6)+\frac{368}{2835}(h_x^4 h_z^2+h_x^2 h_z^4) \nonumber\\
&& -\frac{500690327}{42865200}(h_x^8+h_z^8)+\frac{2420353}{1530900}(h_x^6 h_z^2+h_x^2 h_z^6)+\frac{289031}{595350} h_x^4 h_z^4-\frac{74305313819}{562605750}(h_x^{10}+h_z^{10}) \nonumber\\
&& +\frac{2914018042687}{148527918000}(h_x^8 h_z^2+h_x^2 h_z^8)+\frac{1810666352617}{297055836000}(h_x^6 h_z^4+h_x^4 h_z^6) \quad.\nonumber
\eea

\bea \label{gap-small_paral}
\Delta^{\rm lf} &=& 1-12 h_x^2+32 h_x^4+\frac{46}{9} h_x^2 h_z^2-\frac{134356}{81}h_x^6 -\frac{86054}{2835} h_x^4 h_z^2+\frac{74264}{2835} h_x^2 h_z^4+\frac{57057168481}{2679075}h_x^8\nonumber\\
&& +\frac{663345653}{297675} h_x^6 h_z^2-\frac{134334119}{595350} h_x^4 h_z^4+\frac{625317677}{1786050} h_x^2 h_z^6 \quad.\nonumber
\eea

Ground-state energy per site $\varepsilon^{\rm hf}_{0}$ and 1-QP gap $\Delta^{\rm hf}$ of the TCC in parallel magnetic field in the high-field limit and on the isotropic line ($h_x=h_z=h$) for $h=1/2$:
\bea \label{gs-large_paral}
\varepsilon^{\rm hf}_{0} &=& -\frac{1}{2}-\frac{1}{8} J-\frac{173}{384} J^2+\frac{99}{1024} J^3-\frac{58717}{884736} J^4+\frac{342121}{3538944} J^5 \quad.\nonumber
\eea

\bea\label{gap-large_paral}
\Delta^{\rm hf} &=& 1-\frac{3}{4}J+\frac{137}{64} J^2+\frac{969}{512} J^3+\frac{199337}{442368} J^4+\frac{7492921}{42467328} J^5 \quad.\nonumber
\eea

\section{Series expansion of the TCC in transverse magnetic field}

Ground-state energy per site $\varepsilon^{\rm lf}_{0}$ and 1-, 2- and 3-QP gap of the TCC in transverse magnetic field in the low-field limit for $J=1/2$ (see also Fig.~\ref{bound_state}):
\bea \label{gs-small_trans}
\varepsilon^{\rm lf}_{0} &=& -\frac{1}{2}-\frac{1}{6}h_y^2-\frac{19}{432} h_y^4-\frac{5359}{68040} h_y^6-\frac{500690327}{2743372800} h_y^8-\frac{74305313819}{144027072000} h_y^{10} \quad. \nonumber
\eea

\bea \label{gap-small_trans_1qp}
\Delta^{\rm lf}_{\rm 1QP} &=& 1-\frac{1}{2}h_y^2-\frac{151}{288} h_y^4-\frac{192287}{103680} h_y^6-\frac{1684602101501}{256048128000} h_y^8 \quad.\nonumber
\eea

\bea \label{gap-small_trans_2qp}
\Delta^{\rm lf}_{\rm 2QP} &=& 2-6 h_y^2+4 h_y^4-\frac{33589}{648} h_y^6+\frac{4673722313}{28576800} h_y^8-\frac{29786981411535707}{20739898368000} h_y^{10} \quad.\nonumber
\eea

\bea \label{gap-small_trans_3qp}
\Delta^{\rm lf}_{\rm 3QP} &=& 3-h_y-\frac{19}{12} h_y^2-\frac{3083}{1728} h_y^4-3 h_y^5-\frac{2410397}{435456} h_y^6-\frac{653}{72} h_y^7-\frac{706644011957}{30725775360} h_y^8-\frac{4000507}{294912} h_y^9 \quad.\nonumber
\eea

Ground-state energy per site $\varepsilon^{\rm hf}_{0}$ and 1-QP gap $\Delta^{\rm hf}$ of the TCC in transverse magnetic field in the high-field limit for $h_y=1/2$:
\bea \label{gs-large_trans}
\varepsilon^{\rm hf}_{0} &=& -\frac{1}{2} \quad.\nonumber
\eea

\bea\label{gap-large_trans}
\Delta^{\rm hf} &=& 1-3 J^2+\frac{3}{4} J^4-\frac{15}{32} J^6+\frac{231}{512} J^8-\frac{4227}{8192} J^{10}+\frac{81879}{131072} J^{12}-\frac{1627899}{2097152} J^{14}+\frac{33064431}{33554432} J^{16} \quad.\nonumber
\eea

 \end{document}